\documentclass[a4paper,twoside]{article}
\usepackage{rotating}
\usepackage{multirow}
\newcommand{\affil}[1]{$^{\rm #1}$}

\newcommand{\aap}{A\&A}
\newcommand{\aj}{AJ}
\newcommand{\apj}{ApJ}
\newcommand{\apjl}{ApJL}
\newcommand{\apss}{ApSS}
\newcommand{\araa}{ARA\&A}
\newcommand{\mnras}{MNRAS}
\newcommand{\nat}{Nature}
\newcommand{\pasp}{PASP}
\usepackage{amsfonts}
\usepackage{amssymb}
\usepackage[authoryear]{natbib}
\bibpunct{(}{)}{;}{a}{}{,}
\usepackage{graphicx}
\date{} 
\title{\large\bf\flushleft Resolving the nucleus of Centaurus A at mid-IR wavelengths}
\author{\parbox{\textwidth}{\flushleft
\vspace{-0.5cm}
{\it Leonard Burtscher\affil{A,B,C}, Klaus Meisenheimer\affil{A}, Walter Jaffe\affil{D}, Konrad R. W. Tristram\affil{E} and Huub J. A. R\"ottgering\affil{D}}\\
\vspace{0.4cm}
{\small \affil{A}\,Max-Planck-Institut f\"ur Astronomie, K\"onigstuhl 17, 69217 Heidelberg, Germany}\\
{\small \affil{B}\,Fellow of the International Max Planck Research School (IMPRS) for Astronomy and Cosmic Physics, Heidelberg, Germany}\\
{\small \affil{C}\,Email: burtscher@mpia.de}\\
{\small \affil{D}\,Sterrewacht Leiden, Leiden University, Niels-Bohr-Weg 2, 2300 CA Leiden, The Netherlands}\\
{\small \affil{E}\,Max-Planck-Institut f\"ur Radioastronomie, Auf dem H\"ugel 69, 53121 Bonn, Germany}}}
\begin{document}
\twocolumn[
\begin{changemargin}{.8cm}{.5cm}
\begin{minipage}{.9\textwidth}
\vspace{-1cm}
\maketitle
%
%
\small{\bf Abstract: We have observed Centaurus A with the MID-infrared Interferometric instrument (MIDI) at the Very Large Telescope Interferometer (VLTI) at resolutions of 7 -- 15 mas (at 12.5 ${\bf\mu}$m) and filled gaps in the (u,v) coverage in comparison to earlier measurements. We are now able to describe the nuclear emission in terms of geometric components and derive their parameters by fitting models to the interferometric data. With simple geometrical models, the best fit is achieved for an elongated disk with flat intensity profile with diameter ${\bf (76 \pm 9) mas \times (35 \pm 2) mas}$ (${\bf (1.41 \pm 0.17) pc \times (0.65 \pm 0.03) pc}$) whose major axis is oriented at a position angle (PA) of ${\bf (10.1 \pm 2.2)^{\circ}}$ east of north. A point source contributes ${\bf (47 \pm 11)}$ \% of the nuclear emission at 12.5 ${\bf \mu}$m. There is also evidence that neither such a uniform nor a Gaussian disk are good fits to the data. This indicates that we are resolving more complicated small-scale structure in AGNs with MIDI, as has been seen in Seyfert galaxies previously observed with MIDI.
The PA and inferred inclination ${\bf i = 62.6^{+2.1}_{-2.6} ~ ^{\circ}}$ of the dust emission are compared with observations of gas and dust at larger scales.}


\medskip
\medskip
\end{minipage}
\end{changemargin}
\vspace{0.7cm}
]
\small

\section{Introduction}

In the unified model of Active Galactic Nuclei (AGNs) \citep{antonucci1993}, the central engine (black hole + accretion disk + broad line region clouds) is embedded in a dusty obscuring torus. Galaxies where we see the center only through this torus are called ``type 2'' whereas ones where we receive direct radiation from the broad line region are referred to as ``type 1''. For a long time, there was only indirect evidence for the existence of the enshrouding dust, until VLTI observations with MIDI actually resolved the parsec-scale AGN heated dust structures in NGC 1068 \citep{jaffe2004}. Now, ``tori'' have been observed interferometrically in a number of Seyfert 1 and 2 galaxies \citep{tristram2007b,raban2009,tristram2009,burtscher2009}.

Radio galaxies can also be classified in type 1 (e.g. Broad Line Radio Galaxies, BLRGs) and type 2 (e.g. Narrow Line Radio Galaxies, NLRGs) sources and are also unified by means of orientation dependent obscuration, although a ``torus'' is not required in all cases \citep{urry1995}.

Centaurus A (NGC 5128) is the nearest radio powerful AGN with a distance of only $3.80 \pm 0.08$~Mpc (see Harris et al. in this volume). At this distance, one parsec corresponds to 54 milli-arcseconds (mas). Centaurus A is therefore the only radio galaxy (classification: NLRG) that can be studied with MIDI in great detail.

With limited $(u,v)$ coverage, \citet{meisenheimer2007} found from earlier MIDI data that the mid-IR emission from the central parsec of Centaurus A is dominated (60\% at at 13 $\mu$m) by an unresolved point source (diameter $<$ 10 mas). By comparison with multi-wavelength data, this emission was classified as being most likely non-thermal (synchrotron) in origin. The resolved emission was estimated to be $\gtrsim$ 30 mas (0.6 pc) and $\lesssim$ 12 mas (0.2 pc) in size, roughly perpendicular and along the axis of the radio jet respectively with the major axis at about $127 \pm 9 ~ ^{\circ}$, roughly perpendicular to the radio jet. It was interpreted as a geometrically thin, inclined dusty disk. With only 4 $(u,v)$ points, the PA of the inferred disk was simply fixed to the direction orthogonal to the PA of the maximum visibilities.

Here we report first results from a more extensive set of $(u,v)$ points that allows to fit model source brightness distributions to the nuclear mid-infrared emission and determine the parameters of the possible emission components more precisely. Further analyses regarding the structure of the emitter will be published elsewhere.

\section{Instrument, observations and data reduction}
\label{sec:red}

Observations were performed in the \emph{N} band ($8 \mu\mathrm{m} < \lambda < 13 \mu\mathrm{m}$) with MIDI \citep{leinert2003b} at the ESO VLTI on Cerro Paranal, Chile, using pairs of 8.2\,m Unit Telescopes (UTs). The observables with MIDI are the single-dish spectra and the correlated flux spectrum that is obtained from the interference pattern generated by the two beams. MIDI observations do not provide visibility phases since the phase information is destroyed through atmospheric turbulence. Since MIDI can only combine the light of two telescopes, closure phases are also not available.

The light was dispersed with a NaCl prism with spectral resolution $R\equiv \lambda/\Delta\lambda \sim 30$ and a slit width of 200-$\mu$m that corresponds to 0.52 arcsec on the sky. Since Cen A appears as a point source in the mid-infrared with the single-telescope observations, the effective ``field of view'' of the interferometer observations corresponds to the width of the point-spread function at the wavelength of observation (380 mas, or 7 pc, at 12.5 $\mu$m).

\begin{figure}[t]
	\begin{centering}
		\includegraphics[width=7.5cm, trim = 0cm 1cm 0cm 0cm]{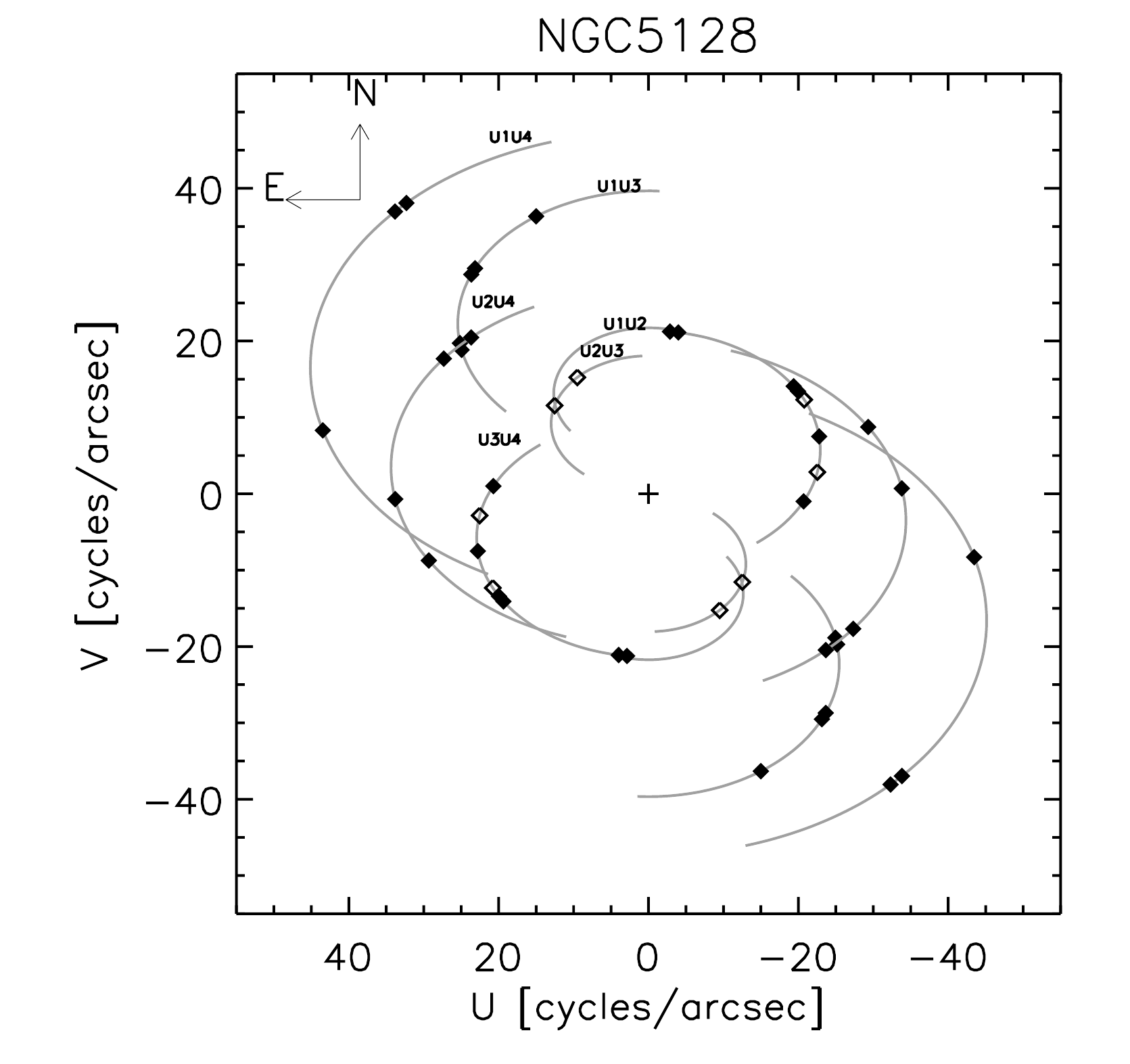}
		\caption{$(u,v)$ coverage of the 2008 MIDI visibilities of Cen A at 12.5 $\mu m$.  Observations that led to \citet{meisenheimer2007} (4 visibilities) are shown as open diamonds, the 2008 observations (18 visibilities) as filled diamonds. Every visibility appears twice in this plot. The solid lines are the $(u,v)$ positions traced by the various baselines (UT combinations are labelled) as a result of earth's rotation. They are truncated at telescope elevations of 30 degrees above the horizon.}
		\label{fig:uv}
	\end{centering}
\end{figure}

For the observation log as well as the observational parameters, see Tab.~\ref{tab:obs}. In total, 20 fringe track observations were made of which one was discarded due to its very low signal. For another one no single-dish observation was taken, leading to a total of 18 visibilities. Some of them (6) are practically duplicates in the $(u,v)$ plane and were taken to increase signal-to-noise. The resulting $(u,v)$ coverage is displayed in Fig.~\ref{fig:uv}. These observations provide effective spatial resolutions\footnote{Applying Rayleigh's criterion to an interferometer leads to a resolution of $\lambda / 2 BL$. However, given sufficient S/N, one can already distinguish models at lower resolutions, in our case at $\sim \lambda / 3 BL$.} at 12.5$\mu$m\, of 6.7 - 15.5 mas, see also Tab.~\ref{tab:obs}. Calibrators were selected to be close in airmass. Due to the orientation of the VLTI baselines and the declination of the source, there are unfortunately no long UT baselines in the lower left (SE) quadrant of the $(u,v)$ plane.

Data reduction was performed with the interferometric data reduction software \emph{MIDI Interactive Analysis and Expert Work Station} \citep[MIA+EWS,][]{jaffe2004b}\footnote{The nightly build from Feb 07 2010 was used. It can be downloaded from http://www.strw.leidenuniv.nl/$\sim$koehler/MIA+EWS-daily-snapshot.html} and the data reduction procedure was as follows: First the associated calibrator (see Tab.~\ref{tab:obs}) was reduced. For the extraction of the spectra, a 4-pixel wide Gaussian-shaped mask was used. The vertical shift of the mask was determined by cross-correlating the mask with the spectrum. Additionally, a skymask was used to determine the sky background at equal, short distances above and below the spectrum.

With these masks, the science data were reduced with identical settings and calibrated using the calibrator star. The calibrated correlated fluxes were determined directly from the raw correlated fluxes of the science and calibrator sources without using the single-dish spectra. This substantially lowers the measurement error for weak sources since the uncertainty of the single-dish fluxes is much larger than that of the correlated fluxes. The intrinsic calibrator spectra were taken from \citet{cohen1999}. A more detailed description of the general MIDI data reduction procedure can be found in \citet{tristram2007}.

All Cen A single-dish spectra were stacked and combined to one average spectrum (for all observations). This is possible since only the correlated spectrum depends on the projected baseline length and orientation and it is not expected that the N band source changes within one week. The 2005 observations that led to \citet{meisenheimer2007} were not included in this analysis since their total flux differs considerably from the 2008 observations.

Finally, the calibrated correlated spectra were divided by the average single-dish spectrum to derive visibilities. For this work, the visibilities between 12.3$\mu$m and 12.7$\mu$m were averaged and all model fits in this paper refer to this part of the N band spectrum. The full spectral information will be exploited in a subsequent paper. The error of the calibrated visibility is composed of the statistical error of the correlated flux and that of the averaged photometry as well as the error of the calibration template (assumed to be 5 \%). The errors of the averaged 12.5 $\mu$m visibilities are then on the order of 3 \%, see also Tab.~\ref{tab:obs}.


\begin{sidewaystable*}
\caption{Main properties of all 2008 NGC 5128 MIDI observations and related calibrators. Observation id, projected baseline length $BL$ and position angle $PA$, effective spatial resolution of the interferometer $\Theta_{\rm min}$, averaged visibility $V$ at $(12.5 \pm 0.2) \mu$m, Airmass and Seeing}
\label{tab:obs}
\centering
\begin{tabular}{c c c c c c c c c c}
\hline\hline
Date and Time & id  & BL  & PA           & $\Theta_{\rm min}$ & $V$ & Airmass & Seeing & Comment / associated calib\\
\textrm{[UTC]}    &   &[m]&[$^{\circ}$]       & [mas (pc)]   &          &         &['']&\\
\hline
2008-04-16: & UT1 -- UT3 \\
\hline
04:05:48 & s1 & 99.55 &        31.29 & &   &     1.06 & 2.89 &         very low signal; not used \\
\hline
\hline
2008-04-18: & UT1 -- UT3\\
\hline
02:54:31 & s1 & 101.35 &        22.43 & 8.5 (0.16)  & $      0.68 \pm       0.04$  &     1.10 & 1.07 &         HD112213 \\
04:51:38 & s2 &96.72 &        38.11 & 8.9 (0.16)  & $      0.61 \pm       0.03$  &     1.06 & 1.05 &          HD112213\\
05:03:06 & s3 &95.92 &        39.48  & 9.0 (0.17)  & $      0.60 \pm       0.03$  &     1.07 & 1.10 &         HD112213\\
07:03:41 & s4  & 82.44 &        51.92 &10.4 (0.19)  & $      0.49 \pm       0.03$  &     1.28 & 1.71 &         HD119193 \\
07:15:19 & s5  & 80.61 &        52.92 & 10.7 (0.20)  & $      0.49 \pm       0.03$  &     1.32 & 1.22 &        HD119193 \\
09:26:26 & s6  &  53.75 &        60.81 &  &   &     2.33 & 1.90 &        very low signal; not used \\
\hline
\hline
2008-04-19: & UT1 -- UT4\\
\hline
02:09:13 & s2  &128.74 &        40.33 & 6.7 (0.12)  & $      0.50 \pm       0.03$  &     1.17 & 0.95 &         HD112213 \\
02:20:33 & s3  &129.17 &        42.48 & 6.7 (0.12)  & $      0.49 \pm       0.02$  &     1.15 & 1.10 &         HD112213  \\
06:08:06 & s8  &  114.07 &        79.20 &7.5 (0.14)  & $      0.38 \pm       0.02$  &     1.15 & 1.54 &     treated as one datapoint with s9; HD112213 \\
06:11:55 & s9  &  113.31 &        79.78 &  7.5 (0.14)  & $      0.38 \pm       0.02$  &     1.16 & 1.23 &      treated as one datapoint with s8; HD112213 \\
\hline
\hline
2008-04-21: & UT3 -- UT4\\
\hline
01:47:30 & s1  & 53.46 &        87.27 & 16.1 (0.30)  & $      0.62 \pm       0.03$  &     1.20 & 0.66 &        HD111915 \\
03:55:10 & s4  & 61.85 &        108.28 &13.9 (0.26)  & $      0.61 \pm       0.03$  &     1.06 & 0.65 &         HD112213 \\
05:26:40 & s5  & 62.02 &        123.96 &13.9 (0.26)  & $      0.58 \pm       0.03$  &     1.10 & 0.60 &         HD112213 \\
05:38:02 & s6  & 61.77 &        126.06 &13.9 (0.26)  & $      0.57 \pm       0.03$  &     1.12 & 0.68 &         HD112213 \\
08:48:03 & s9  & 55.44 &        169.34 &15.5 (0.29)  & $      0.47 \pm       0.02$  &     1.98 & 0.57 &         HD119193 \\
08:59:20 & s10  & 55.27 &        172.34 & 15.5 (0.29)  & $      0.53 \pm       0.03$  &     2.12 & 0.54 &        HD119193 \\
\hline
\hline
2008-04-23: & UT2 -- UT4\\
\hline
01:06:08 & s1  & 80.68 &        49.15 &10.7 (0.20)  & $      0.47 \pm       0.02$  &     1.30 & 1.17 &         HD119193 \\
01:42:35 & s2  & 83.92 &        57.08 & 10.2 (0.19)  & $      0.56 \pm       0.03$  &     1.20 & 0.69 &        HD111915 \\
04:52:34 & s8  & 87.18 &        91.22 &9.9 (0.18)  & $      0.54 \pm       0.03$  &     1.08 & 1.00 &         HD112213 \\
06:16:51 & s9  & 78.88 &        106.62 &10.9 (0.20)  & $      0.70 \pm       0.04$  &     1.21 & 1.08 &         HD119193 \\
\hline
\end{tabular}
\end{sidewaystable*}

%

\section{Results and modelling}
\label{sec:results}

\subsection{Visibilities}
\begin{figure}[t]
	\begin{centering}
		\includegraphics[width=6.5cm, trim = 1cm 1cm 1cm 1cm]{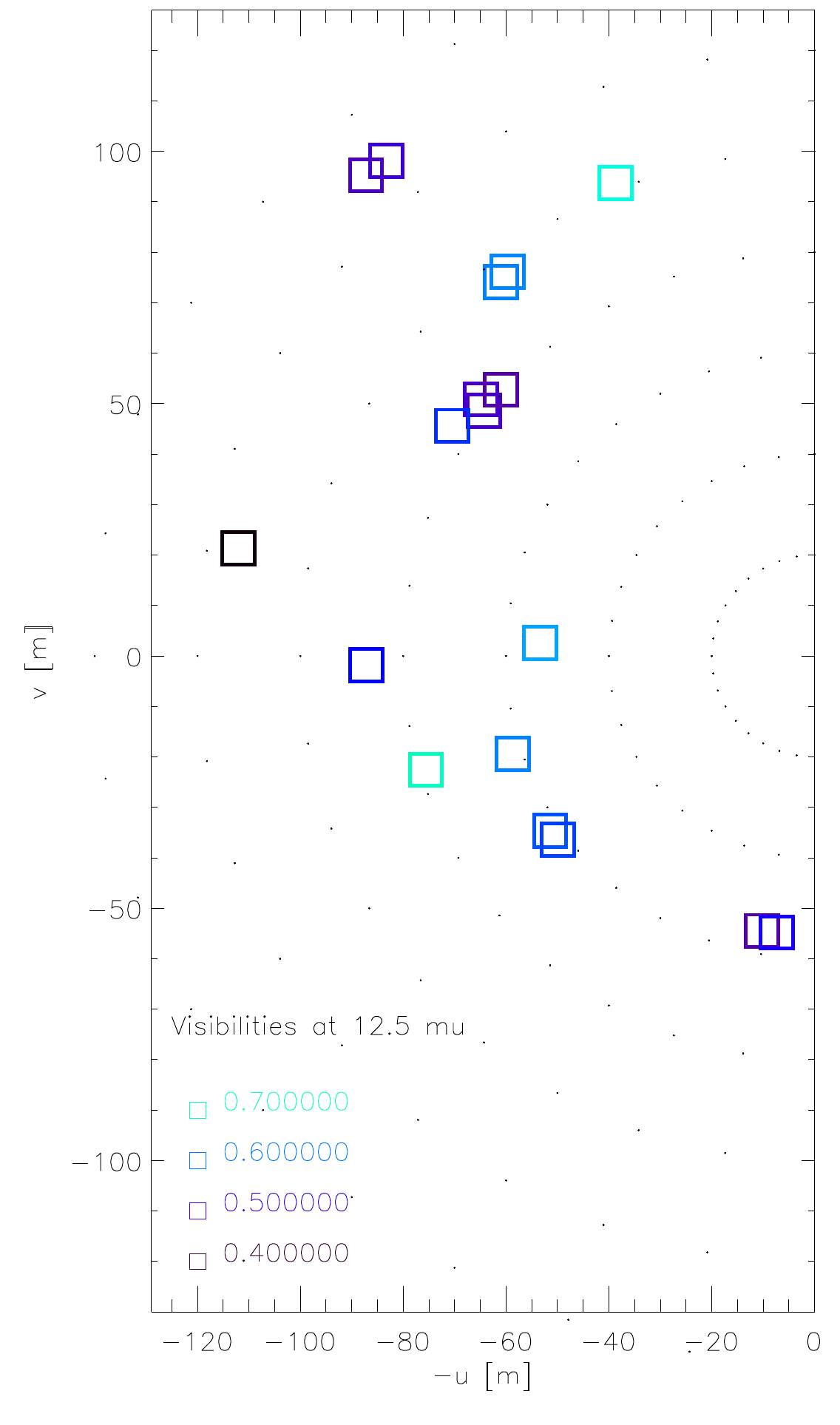}
		\caption{Visibilities in the $(u,v)$ plane. Each symbol represents a visibility measurement, the color of the symbol shows the visibility amplitude. The dots denote radii of 20, 40 ... 140~m around $(u,v)$ = 0. The data from \citet{meisenheimer2007} are not shown in this plot.}
		\label{fig:uv-vis}
	\end{centering}
\end{figure}

The resulting averaged visibilities at 12.5 $\mu$m ($\Delta\lambda = 0.2\mu$m), measured at baselines of about 50 -- 130 meters, range from about 40\% -- 70\%. They are given in Tab.~\ref{tab:obs} and visualised in a $(u,v)$ plot in Fig.~\ref{fig:uv-vis} where each symbol represents a fringe observation. The color of the symbol denotes the visibility. Their distribution in the $(u,v)$ plane does not intuitively show what structure they probe.

%

\subsection{Geometric models for the surface brightness distribution}

With phase-less data on sparsely sampled $(u,v)$ planes (see Fig.~\ref{fig:uv}), it is impossible to directly reconstruct meaningful images, i.e. images that show more than just the properties of the synthesized beam. We therefore model the source brightness distribution (the ``image'') with simple geometrical components (e.g. point sources, rings, ellipses etc.) and constrain the parameters of the model with the observed visibilities \citep[e.g.][]{berger2007}.

Here we present our model fit to the visibilities at one wavelength. The fit was performed with Lyon's Interferometric Tool prototype (LITpro), an interferometric model fitting tool, provided by the Jean-Marie Mariotti Center \citep[JMMC,][]{tallon-bosc2008}\footnote{The online version of LITpro can be found at http://www.jmmc.fr/litpro\_page.htm}. It is based on a modified Levenberg-Marquardt algorithm and runs a grid-search to find the least $\chi^2$ value in the parameter space (in given boundaries).

\subsubsection{Two-components models}

Motivated by the finding of \citet{meisenheimer2007} that the infrared nuclear emission is composed of a resolved and an unresolved component, we fitted two-component models with a point source (unresolved) and an extended component (resolved). We model the extended source either with a circularly symmetric Gaussian intensity profile (model {\em a}), an elongated Gaussian intensity profile (model {\em b}) or with an elongated uniform (flat intensity profile) disk (model {\em c}). A circularly symmetric flat profile component, a ring-shaped emission component as well as limb-darkened profiles lead to unphysical parameters.

All models are characterized by the flux fraction of the point source ($f_p$).

The size parameter $\Theta$ has model-specific meanings:

\begin{itemize}
	\item[\emph{model a}] Gaussian Full-Width at Half Maximum (FWHM)
	\item[\emph{model b}] minor axis FWHM
	\item[\emph{model c}] minor axis diameter
\end{itemize}

The models with an elongated extended component ({\em b}, {\em c}) additionally have the parameters axis ratio $\rho$ and major axis position angle $PA$ in degrees east of north.

The center positions of the two components were held fixed on top of each other. Models with non-concentric components will be discussed in a future paper.

\begin{figure}
	\begin{centering}
		\includegraphics[width=6cm, trim = 1cm 0cm 1cm 1cm]{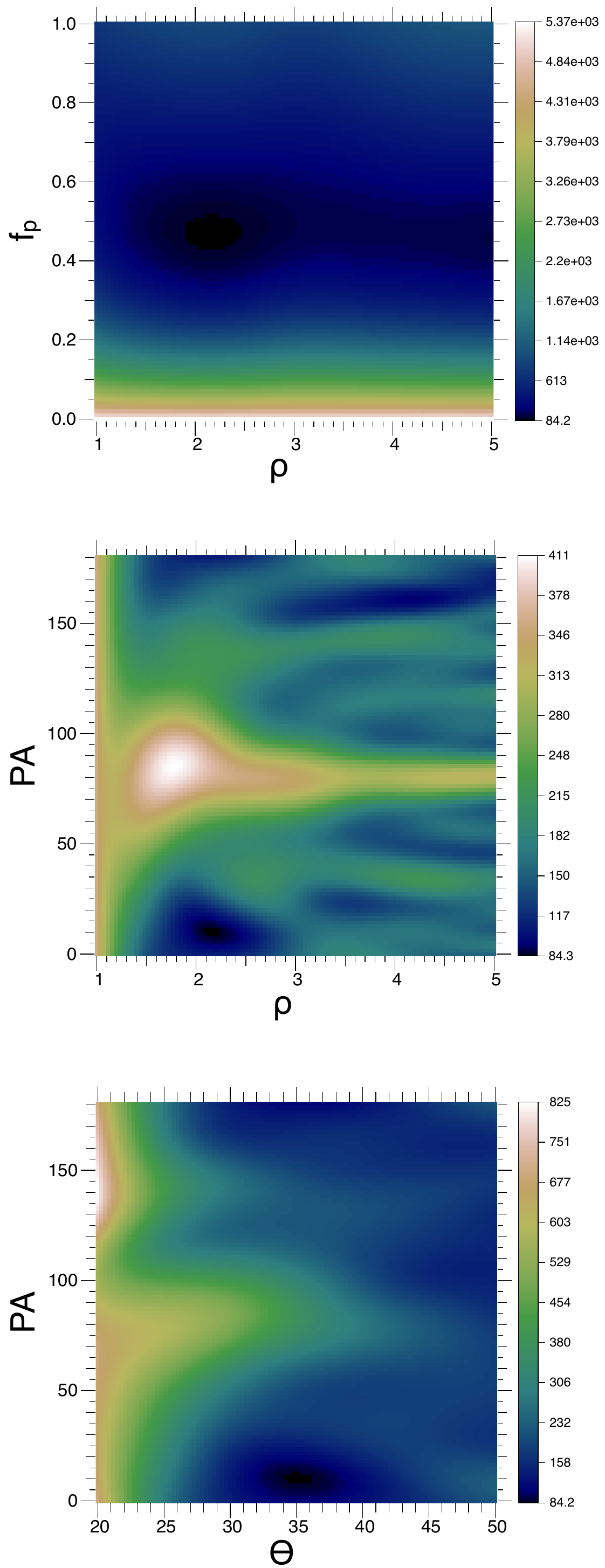}
		\caption{$\chi^2$ cuts of our model {\emph c} (elongated disk). The color denotes the value of $\chi^2$ (white: maximum; black: minimum, see color wedge). {\em Top:} flux fraction of the point source versus axis ratio $\rho$; {\em Middle:} major axis position angle versus axis ratio $\rho$; {\em Bottom:} major axis position angle versus minor axis diameter $\Theta$. See text for discussion.}
		 \label{fig:x2-combined}
	\end{centering}
\end{figure}

\begin{figure}[t]
	\begin{centering}
		\includegraphics[width=6.5cm, trim = 1cm 1cm 1cm 1cm]{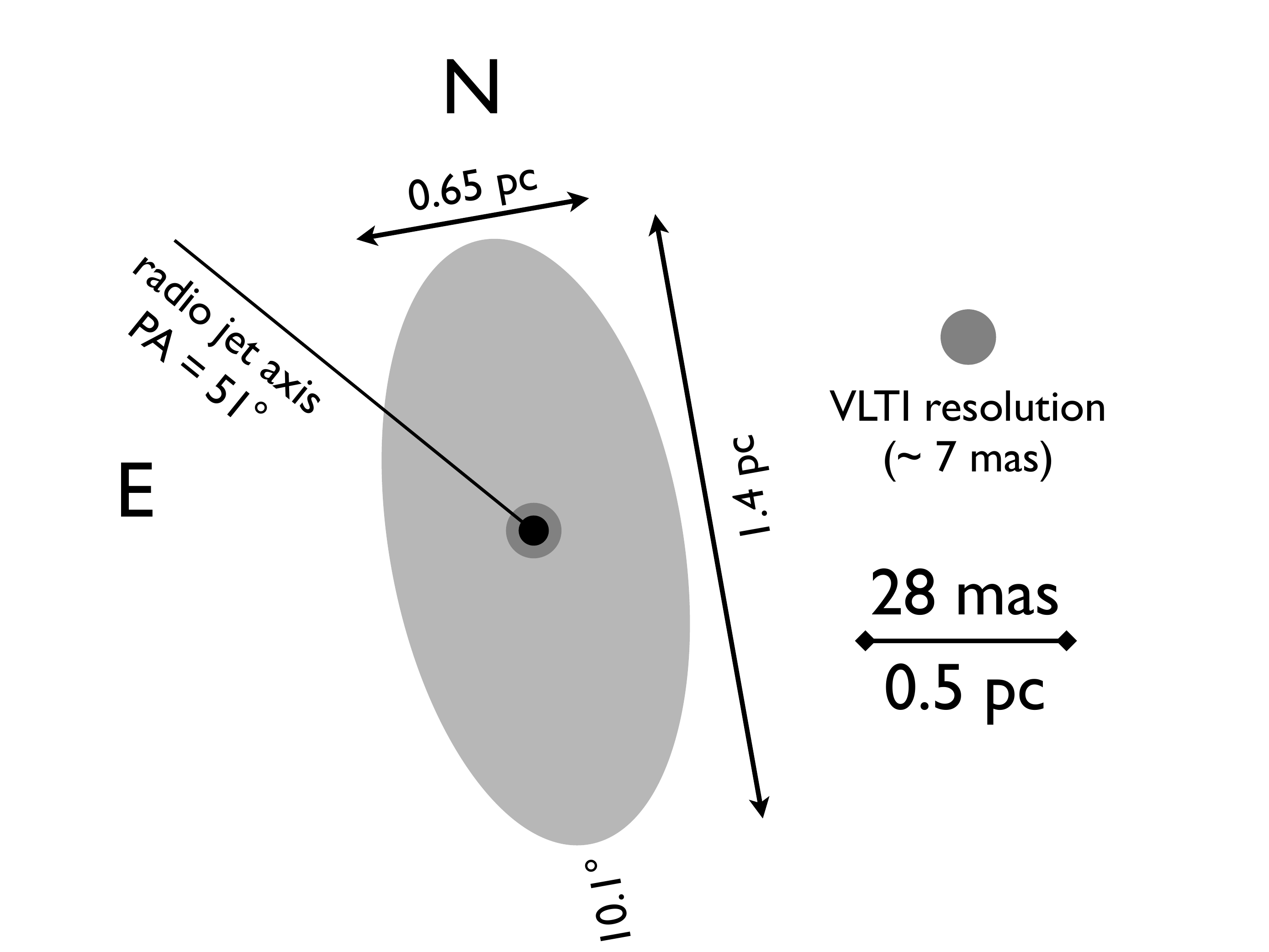}
		\caption{Two-component model for the nuclear mid-infrared emission of Centaurus~A consisting of a uniform disk component and an unresolved point source that is smaller than the VLTI resolution (dark grey circle) and has been identified with the synchrotron core (black point) by \citet{meisenheimer2007}. The resolved structure is $\sim 1.4 \times 0.65$ pc at 12.5 $\mu$m and the best fit PA $\sim 10.1^{\circ}$; the radio radio jet axis is PA $\sim 51^{\circ}$ \citep[e.g.][]{tingay2001}.}
		\label{fig:model}
	\end{centering}
\end{figure}

The best fit (in terms of $\chi^2$) can be achieved by fitting the resolved component as an elongated disk of $(1.41 \pm 0.17) \times (0.65 \pm 0.03)$ pc. The major axis of this structure is almost north--south at a PA of $(10.1 \pm 2.2)^{\circ}$. In this model, the point source contributes $(47 \pm 11)$ \% to the total flux at 12.5 $\mu$m. The parameters of all models are compiled in Tab.~\ref{tab:fitresults}. Fig.~\ref{fig:x2-combined} shows cuts through the $\chi^2$ plane of the best-fitting model ({\em c}) to show how well constrained the parameters are: While the minimum for the point source flux contribution is well defined, the $\chi^2$ planes show that PA and axis ratio $\rho$ are not so well defined. The $\chi^2$ plane for PA and $\rho$ shows a number of local minima, indicating degeneracies in the model fit. Fig.~\ref{fig:model} shows a sketch of this model.

%

\begin{table}
\caption{Two-component model fits. For the description of the parameters, see text. For each model, the $\chi^2$ value for the best fitting parameters is given together with the number of degrees of freedom (NDOF).}
\begin{center}
\begin{tabular}{|c | c c c|}
\hline
                     & a: Gauss          & b: El. Gauss          & c: El. disk \\ \hline
$f_p$                & $0.48 \pm 0.11$   & $0.49 \pm 0.12$  & $0.47 \pm 0.11$ \\ \hline
$\Theta$ / mas       & $38.7 \pm 3.0$    & $22.9 \pm 1.6$   & $35.2 \pm 1.5$ \\ 
$\Theta$ / pc        & $0.72 \pm 0.06$   & $0.42 \pm 0.03$  & $0.65 \pm 0.03$ \\ \hline
$\rho$               &   ---             & $2.94 \pm 0.53$  & $2.17 \pm 0.17$ \\ \hline
$PA / ^{\circ}$      &   ---             & $14.9 \pm 2.4$   & $10.1 \pm 2.2$ \\ \hline
NDOF                 &   15              &   13             &  13      \\
$\chi^2$             &  {\bf 135.0}      &  {\bf 97.6}      & {\bf 84.1}   \\
\hline
\end{tabular}
\end{center}
\label{tab:fitresults}
\end{table}%

%

\section{Discussion}

The parameters of our disk are different from the \citet{meisenheimer2007} result: They find a $\gtrsim 0.6 {\rm pc} \times \lesssim 0.2 {\rm pc}$ Gaussian disk at PA $\sim 127^{\circ}$. The probable origin for this difference is their sparse $(u,v)$ coverage. Had we only had two pairs of observations, e.g. s1/s2 of 2008-04-18 and s9/s10 of 2008-04-21, our results would be closer to \citet{meisenheimer2007}: Converting the visibilities of these observations to one-dimensional Gaussian widths, leads to full-widths at half-maxima of roughly 10 mas (0.2 pc) along PA $\approx 30^{\circ}$ and 20 mas (0.4 pc) along PA $\approx 170^{\circ}$. This shows how important it is to study the circumnuclear dusty component of AGNs not only with a few baselines but with many different telescope combinations.

In fact, the visibilities look rather chaotic with values as high as 70\% separated by only about two telescope diameters from values as low as 55\%. Possible causes for these variations, other than the effects from the science source itself, could be variations of the atmospheric transfer function between calibrator and science target. Typical variations of the transfer function were $\approx 3\%$ in between science and calibrator observation as determined by linear interpolation of the transfer function between each two subsequent calibrator observations.
Another possibility for artificial flux variations are differences in the beam profile between observations (e.g. due to vignetting). This effect could be important since we did not divide each correlated flux spectrum by its corresponding single-dish spectrum individually but averaged all single-dish spectra in order to reduce noise (see Section \ref{sec:red}). We tested this by looking for differences in the calibrator single-dish beam profiles and found none of significance.
Furthermore, it is possible that spectra observed on different baselines have systematic offsets. We tested this by observing the same (u,v) point twice on two different baselines. The scatter between the baselines was not larger than the scatter within the baseline.

These possible causes for artificial visibility variations cannot explain the differences observed. Therefore we believe these offsets are real.

In other sources such jumps in the visibility are only seen at very low visibility levels ($\lesssim 20$ \%) where the source is almost completely resolved (cf. Fig. 4 of \citet{tristram2007b}). But in Cen A the point source contributes almost 50\% to the total flux according to our model fits. Therefore, we are effectively observing the extended component at visibilities $\lesssim 20$ \%. This means we are probing the extended component at relatively high spatial resolutions which are most sensitive to small-scale structures and not well described by a Gaussian or uniform disk. 

This is not only seen in our Cen A data, where our best fit model results in a $\chi^2 = 84.1$ (with 13 degrees of freedom), but also in the two other AGNs that have been studied extensively with MIDI: The Circinus galaxy and NGC 1068. In the Circinus galaxy, a fit with two Gaussian components leads to a $\chi^2 \sim 16 600$ with 451 degrees of freedom \citep{tristram2007b}; in NGC 1068 especially the data at longer baselines are also not well described by a model of two Gaussian components \citep{raban2009}. These small scale structures could, for example, arise from clumpiness in the dusty disk as has been demonstrated by \citet{tristram2007b}.

It is therefore not completely unexpected that the geometric fit to the Cen A data doesn't fully describe the visibility data.

A comparison of our models {\em a}, {\em b}, {\em c} shows that (i) an elongated disk considerably improves the fit and (ii) a Gaussian profile leads to a similar disk area ($\sim 67 {\rm mas} \times 23 {\rm mas}$, corresponding to $\sim$ 1.2 pc $\times$ 0.4 pc) as a flat profile ($\sim 76 {\rm mas} \times 35 {\rm mas}$, corresponding to $\sim$ 1.4 $\times$ 0.6 pc) at very similar position angles of $(14.9 \pm 2.4)^{\circ}$ and $(10.1 \pm 2.2)^{\circ}$, respectively. However, when interpreting the reliability of the PA of the disk in such fits, the asymmetric $(u,v)$ coverage (Figs. \ref{fig:uv}, \ref{fig:uv-vis}) must be taken into account that could artificially create elongated disks. This can possibly be constrained with future observations.

The highest effective spatial resolutions of the new 12.5 $\mu$m data presented here are still about 10$\times$ larger than the sublimation radius of dust in Cen A \citep{meisenheimer2007} and the point source of our model could therefore still be contaminated by dust emission from the innermost part of the dusty disk. A detailed spectral model, taking into account also the visibilities at other wavelengths, will be able to decompose the two components more precisely.

\subsection{Comparison with other observations}

\citet{neumayer2007} used adaptive optics assisted integral field spectroscopy in the near infrared to derive a model for the mass distribution of the central 3 arcsec (60 pc) from the observation of molecular hydrogen emission. At their resolution limit of 120 mas (ca. 2 pc) in the $K$ band, they find a PA of $(148.5 \pm 1)^{\circ}$ and an inclination of $(37.5 \pm 2) ^{\circ}$ for the velocity field in their tilted ring model. 

\citet{espada2009} observed the central 1 arcmin (1 kpc) of Cen A in the $^{12}{\rm CO}(2-1)$ emission line with the Submillimeter Array with a resolution of 100 pc x 40 pc. Using a warped thin disk model, they find that this molecular gas emission in the inner 20 arcsec (400 pc) is elongated at a PA of $\approx 155^{\circ}$ and they interpret the observed axis-ratio of the presumably circular disk as an inclination of $i = 70^{\circ}$. In VLBI observations, \citet{tingay2001} found a position angle of the jet of $51^{\circ}$ and confirmed the previously found inclination of $50^{\circ} - 80^{\circ}$ \citep{tingay1998}. In combined radio and X-ray observations, \citet{hardcastle2003b} find that the radio and X-ray jet trace each other quite closely in terms of position angle and that only small inclinations (angles to the line of sight) of $\approx 20^{\circ}$ are consistent with both X-rays and radio observations from combined constraints on apparent motions and sidedness of the jet (components). The PA of the X-ray jet is $\sim 55 ^{\circ}$ \citep{kraft2000}.

Assuming that the dusty component is well represented by an inclined disk, our axis ratio of $\rho = 2.17 \pm 0.17$ can be interpreted as an inclination of $\cos^{-1} 1/\rho = 62.6^{+2.1}_{-2.6} ~ ^{\circ}$ and would then lie roughly in between the measurements of \citet{neumayer2007} and \citet{espada2009}.

The position angle of $\sim 10^{\circ} (\equiv 190^{\circ})$ from our fit is rather unexpected when compared with these observations on larger scales, however. From the naive unification picture argument, one might expect the gas and dust to be elongated along $\sim 140^{\circ}$, at a right angle to the radio and X-ray jet. This differs by $\sim 50^{\circ}$ from our model fit. Before explaining this with physical effects (warped accretion disk?), however, we need to better understand the chaotic visibility pattern in our observations to better constrain the position angle of the resolved component.

\section{Summary}
We presented new interferometric data from MIDI/VLTI observations that allow us to constrain the geometry of the mid-infrared emission at 12.5 $\mu$m from the nucleus of Centaurus A. We find that, in the context of simple geometrical models, this emission is best reproduced by a two-component model of an unresolved point-source ($\lesssim$ 7 mas) and an elongated (Gaussian or flat) disk with a size of about $1.3 \times 0.5$ pc (average of models {\em b} and {\em c}). Assuming that the dusty component is well represented by an inclined disk, the disk has a PA of about $12.5^{\circ}$ and is inclined by about $66^{\circ}$ (averages of models {\em b} and {\em c})

The new data have also shown that the extended component is almost completely resolved and that we are probing it at very low visibilities / high spatial resolutions -- and that we probably resolve small scale structure that is not described by our geometrical model. The fit parameters for the disk should therefore be interpreted only as tentative evidence for such a structure.

\section*{Acknowledgements}
This research is based on observations collected at the European Southern Observatory, Chile, programme number 081.B-0121 and has made use of the Jean-Marie Mariotti Center \texttt{LITpro} service co-developped by CRAL, LAOG and FIZEAU. The authors wish to thank the anonymous referee for advice in improving the paper. LB wishes to thank Roy van Boekel, J\"org-Uwe Pott, Christian Leipski, Ren\'e Andrae and Hans-Walter Rix for helpful discussions.

\bibliographystyle{aa}

\begin{thebibliography}{22}
\expandafter\ifx\csname natexlab\endcsname\relax\def\natexlab#1{#1}\fi

\bibitem[{{Antonucci}(1993)}]{antonucci1993}
{Antonucci}, R. 1993, \araa, 31, 473

\bibitem[{{Berger} \& {Segransan}(2007)}]{berger2007}
{Berger}, J.~P. \& {Segransan}, D. 2007, New Astronomy Review, 51, 576

\bibitem[{{Burtscher} {et~al.}(2009){Burtscher}, {Jaffe}, {Raban},
  {Meisenheimer}, {Tristram}, \& {R{\"o}ttgering}}]{burtscher2009}
{Burtscher}, L., {Jaffe}, W., {Raban}, D., {et~al.} 2009, \apjl, 705, L53

\bibitem[{{Cohen} {et~al.}(1999){Cohen}, {Walker}, {Carter}, {Hammersley},
  {Kidger}, \& {Noguchi}}]{cohen1999}
{Cohen}, M., {Walker}, R.~G., {Carter}, B., {et~al.} 1999, \aj, 117, 1864

\bibitem[{{Espada} {et~al.}(2009){Espada}, {Matsushita}, {Peck}, {Henkel},
  {Iono}, {Israel}, {Muller}, {Petitpas}, {Pihlstr{\"o}m}, {Taylor}, \&
  {Dinh-V-Trung}}]{espada2009}
{Espada}, D., {Matsushita}, S., {Peck}, A., {et~al.} 2009, \apj, 695, 116

\bibitem[{{Hardcastle} {et~al.}(2003){Hardcastle}, {Worrall}, {Kraft},
  {Forman}, {Jones}, \& {Murray}}]{hardcastle2003b}
{Hardcastle}, M.~J., {Worrall}, D.~M., {Kraft}, R.~P., {et~al.} 2003, \apj,
  593, 169

\bibitem[{{Jaffe} {et~al.}(2004){Jaffe}, {Meisenheimer}, {R{\"o}ttgering},
  {Leinert}, {Richichi}, {Chesneau}, {Fraix-Burnet}, {Glazenborg-Kluttig},
  {Granato}, {Graser}, {Heijligers}, {K{\"o}hler}, {Malbet}, {Miley},
  {Paresce}, {Pel}, {Perrin}, {Przygodda}, {Schoeller}, {Sol}, {Waters},
  {Weigelt}, {Woillez}, \& {de Zeeuw}}]{jaffe2004}
{Jaffe}, W., {Meisenheimer}, K., {R{\"o}ttgering}, H.~J.~A., {et~al.} 2004,
  \nat, 429, 47

\bibitem[{{Jaffe}(2004)}]{jaffe2004b}
{Jaffe}, W.~J. 2004, in Society of Photo-Optical Instrumentation Engineers
  (SPIE) Conference Series, Vol. 5491, Society of Photo-Optical Instrumentation
  Engineers (SPIE) Conference Series, ed. W.~A. {Traub}, 715--+

\bibitem[{{Kraft} {et~al.}(2000){Kraft}, {Forman}, {Jones}, {Kenter}, {Murray},
  {Aldcroft}, {Elvis}, {Evans}, {Fabbiano}, {Isobe}, {Jerius}, {Karovska},
  {Kim}, {Prestwich}, {Primini}, {Schwartz}, {Schreier}, \&
  {Vikhlinin}}]{kraft2000}
{Kraft}, R.~P., {Forman}, W., {Jones}, C., {et~al.} 2000, \apjl, 531, L9

\bibitem[{{Leinert} {et~al.}(2003){Leinert}, {Graser}, {Przygodda}, {Waters},
  {Perrin}, {Jaffe}, {Lopez}, {Bakker}, {B{\"o}hm}, {Chesneau}, {Cotton},
  {Damstra}, {de Jong}, {Glazenborg-Kluttig}, {Grimm}, {Hanenburg}, {Laun},
  {Lenzen}, {Ligori}, {Mathar}, {Meisner}, {Morel}, {Morr}, {Neumann}, {Pel},
  {Schuller}, {Rohloff}, {Stecklum}, {Storz}, {von der L{\"u}he}, \&
  {Wagner}}]{leinert2003b}
{Leinert}, C., {Graser}, U., {Przygodda}, F., {et~al.} 2003, \apss, 286, 73

\bibitem[{{Meisenheimer} {et~al.}(2007){Meisenheimer}, {Tristram}, {Jaffe},
  {Israel}, {Neumayer}, {Raban}, {R{\"o}ttgering}, {Cotton}, {Graser},
  {Henning}, {Leinert}, {Lopez}, {Perrin}, \& {Prieto}}]{meisenheimer2007}
{Meisenheimer}, K., {Tristram}, K.~R.~W., {Jaffe}, W., {et~al.} 2007, \aap,
  471, 453

\bibitem[{{Neumayer} {et~al.}(2007){Neumayer}, {Cappellari}, {Reunanen}, {Rix},
  {van der Werf}, {de Zeeuw}, \& {Davies}}]{neumayer2007}
{Neumayer}, N., {Cappellari}, M., {Reunanen}, J., {et~al.} 2007, \apj, 671,
  1329

\bibitem[{{Quillen} {et~al.}(2008){Quillen}, {Bland-Hawthorn}, {Green},
  {Smith}, {Prasad}, {Alonso-Herrero}, {Cleary}, {Brookes}, \&
  {Lawrence}}]{quillen2008}
{Quillen}, A.~C., {Bland-Hawthorn}, J., {Green}, J.~D., {et~al.} 2008, \mnras,
  384, 1469

\bibitem[{{Raban} {et~al.}(2009){Raban}, {Jaffe}, {R{\"o}ttgering},
  {Meisenheimer}, \& {Tristram}}]{raban2009}
{Raban}, D., {Jaffe}, W., {R{\"o}ttgering}, H., {Meisenheimer}, K., \&
  {Tristram}, K.~R.~W. 2009, \mnras, 394, 1325

\bibitem[{{Schartmann} {et~al.}(2009){Schartmann}, {Meisenheimer}, {Klahr},
  {Camenzind}, {Wolf}, \& {Henning}}]{schartmann2009}
{Schartmann}, M., {Meisenheimer}, K., {Klahr}, H., {et~al.} 2009, \mnras, 393,
  759

\bibitem[{{Tallon-Bosc} {et~al.}(2008){Tallon-Bosc}, {Tallon}, {Thi{\'e}baut},
  {B{\'e}chet}, {Mella}, {Lafrasse}, {Chesneau}, {Domiciano de Souza},
  {Duvert}, {Mourard}, {Petrov}, \& {Vannier}}]{tallon-bosc2008}
{Tallon-Bosc}, I., {Tallon}, M., {Thi{\'e}baut}, E., {et~al.} 2008, in
  Presented at the Society of Photo-Optical Instrumentation Engineers (SPIE)
  Conference, Vol. 7013, Society of Photo-Optical Instrumentation Engineers
  (SPIE) Conference Series

\bibitem[{{Tingay} {et~al.}(1998){Tingay}, {Jauncey}, {Reynolds}, {Tzioumis},
  {King}, {Preston}, {Jones}, {Murphy}, {Meier}, {van Ommen}, {McCulloch},
  {Ellingsen}, {Costa}, {Edwards}, {Lovell}, {Nicolson}, {Quick}, {Kemball},
  {Migenes}, {Harbison}, {Jones}, {White}, {Gough}, {Ferris}, {Sinclair}, \&
  {Clay}}]{tingay1998}
{Tingay}, S.~J., {Jauncey}, D.~L., {Reynolds}, J.~E., {et~al.} 1998, \aj, 115,
  960

\bibitem[{{Tingay} {et~al.}(2001){Tingay}, {Preston}, \&
  {Jauncey}}]{tingay2001}
{Tingay}, S.~J., {Preston}, R.~A., \& {Jauncey}, D.~L. 2001, \aj, 122, 1697

\bibitem[{{Tristram}(2007)}]{tristram2007}
{Tristram}, K.~R.~W. 2007, PhD thesis, Max-Planck-Institut f{\"u}r Astronomie,
  K{\"o}nigstuhl 17, 69117 Heidelberg, Germany

\bibitem[{{Tristram} {et~al.}(2007){Tristram}, {Meisenheimer}, {Jaffe},
  {Schartmann}, {Rix}, {Leinert}, {Morel}, {Wittkowski}, {R{\"o}ttgering},
  {Perrin}, {Lopez}, {Raban}, {Cotton}, {Graser}, {Paresce}, \&
  {Henning}}]{tristram2007b}
{Tristram}, K.~R.~W., {Meisenheimer}, K., {Jaffe}, W., {et~al.} 2007, \aap,
  474, 837

\bibitem[{{Tristram} {et~al.}(2009){Tristram}, {Raban}, {Meisenheimer},
  {Jaffe}, {R{\"o}ttgering}, {Burtscher}, {Cotton}, {Graser}, {Henning},
  {Leinert}, {Lopez}, {Morel}, {Perrin}, \& {Wittkowski}}]{tristram2009}
{Tristram}, K.~R.~W., {Raban}, D., {Meisenheimer}, K., {et~al.} 2009, \aap,
  502, 67

\bibitem[{{Urry} \& {Padovani}(1995)}]{urry1995}
{Urry}, C.~M. \& {Padovani}, P. 1995, \pasp, 107, 803

\end{thebibliography}

\end{document}